\documentclass{article}

\usepackage{graphicx}  % got figures? uncomment this
\begin{document}
\begin{titlepage}
\vspace*{-2cm}
\flushright{ULB-TH/09-16}\\

\vskip 1.5cm
\begin{center}
{\Large \bf Can colliders disprove leptogenesis?}
\end{center}
\vskip 0.5cm
\begin{center}
{\large Jean-Marie Fr\`ere}\footnote{frere@ulb.ac.be}
\vskip .7cm
Service de Physique Th\'eorique,\\
\vspace{0.5mm}
Universit\'e Libre de Bruxelles, 1050 Brussels, Belgium\\
\end{center}
\vskip 0.5cm
\begin{abstract}
While leptogenesis is a very solid but hard to check contender for the generation of the
observed excess of baryons over anti-baryons in the Universe, we show that the observation
of gauge bosons associated with right-handed currents at present or future colliders would
suffice to disprove its most canonical mechanism.
\end{abstract}

\end{titlepage}
\setcounter{footnote}{0}
\vskip2truecm

\section{Introduction}
This short article will outline a suggestion, not really for testing
leptogenesis in general terms, but rather to disprove it, should some
gauge bosons coupled to the right-handed fermions (we will call them
generically ``Right-Handed W's'' or $W_R$) be discovered at current
or future colliders.

We begin by a quick recapitulation of the leptogenesis scheme, insisting
on its attractiveness, its robustness, but also on the difficulty to submit
it to experimental verification. We will later insist on the fact that
extended gauge symmetries are the natural framework for leptogenesis,
and show that the discovery of right-handed W's would infirm the ``canonical''
leptogenesis mechanism.

Full details of this latter analysis can be found in our common
 work with Thomas Hambye and Gilles Vertongen ~\cite{Frere:2008ct}, where
a more complete bibliography is also provided.

\section{Why leptogenesis?}
The current excess of baryons (in fact we don't know about matter
in general, since we can't count the cosmic background neutrinos) over
anti-baryons is one of the big observational evidences calling for
explanation. A first suggestion came from Grand Unified theories,
more specifically SU(5), but quickly met with an objection related
to the late evolution of the Universe. Anomalies and the resultant
non-conservation of B and L, when operative at the electroweak
transition could indeed destroy  a previously generated baryon asymmetry
on the simple condition that it be consistent with $B-L = 0$, which
is precisely the case in SU(5).

The obvious answers are to use this late occasion  either as a new source of B
generation (electroweak baryogenesis), or as a way to mutate a previously
generated asymmetry into the observed B number: in this latter case, it
is necessary that the previous asymmetry satisfy $ B-L \neq 0$.

As is now well-known, the first possibility, despite its elegance, fails in the
Standard Model alone, for lack both of sufficient CP violation, and of
the out-of-equilibrium component which requires a first order phase transition.
This can be fixed in more extended models (additional singlets, supersymmetry), but
the scheme keeps requiring new CP violation, and depends very heavily on the
poorly controlled dynamics of the B and L violation at the phase transition.

The choice solution therefore has become leptogenesis ~\cite{fy}. In its canonical form,
it is closely associated to the see-saw mechanism, where heavy right-handed
neutrinos coupled by Yukawas to the left-handed ones, are used to generate
the very small observed masses. The large Majorana mass of the neutrinos provides
the necessary L violation, the small Yukawa couplings provide the out-of-equilibrium decays, in such a way that a very robust L asymmetry is generated at high temperature. At the electroweak phase transition, a fraction of this L is converted into a baryonic asymmetry. One of the big advantages is that this conversion process operates by reaching  some equilibrium between B and L components, and is fairly independent on the precise dynamics of the B violation at the (slow) electroweak phase transition.

The difficulty to prove leptogenesis resides precisely in its sturdiness, and its quite generic character. Even if the main elements appearing in the calculation of the leptonic (and later baryonic) asymmetry are the same as those governing the (accessible) light neutrino masses, they intervene in completely different combinations, so that low energy data are not constraining for the process.

In this note, we will show that, even if leptogenesis is difficult to establish,
and fairly resilient as a mechanism, it could still be excluded if $W_R$ particles
are observed at colliders.

\section{Orders of Magnitude}
Let us take as a starting point the mass terms for the heavy right-handed neutrinos $N$, and their Yukawa couplings
to the light ones, namely:

\begin{equation}
{\cal L}_{mass} =
- \overline{L} \,{\widetilde H} \, {\lambda_\nu^\dagger}  \, N
-\frac{1}{2}\,\overline{N} \, {m_N} \,{{N}^c}
+\mbox{h.c.}
\end{equation}

where $\lambda$ is a matrix in generation space, $H$ is the Brout-Englert-Higgs doublet (possibly
part of a larger grand-unified multiplet) , $L$ are the light left-handed fermions.

Since we are just interested here in orders of magnitude, we will use in this paragraph $\lambda$
as a single number, assuming (wrongly) that all Yukawa couplings are of similar size.
We want now to express the conditions (the values of $\lambda$)  that provide the correct order
of magnitude for light neutrino masses, for the out-of-equilibrium decay of the heavy $N$,
and for sufficient CP violation.

\begin{figure}
  \includegraphics[width=4cm]{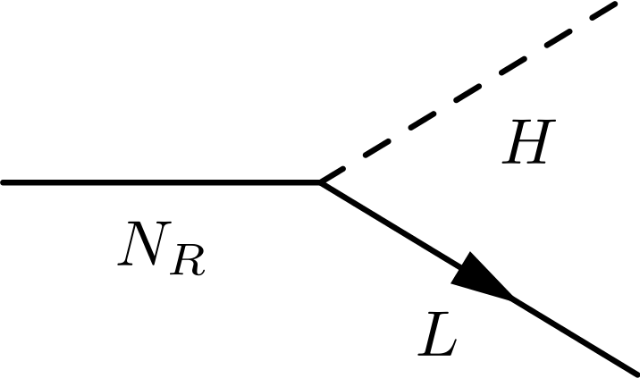}
  \includegraphics[width=4cm]{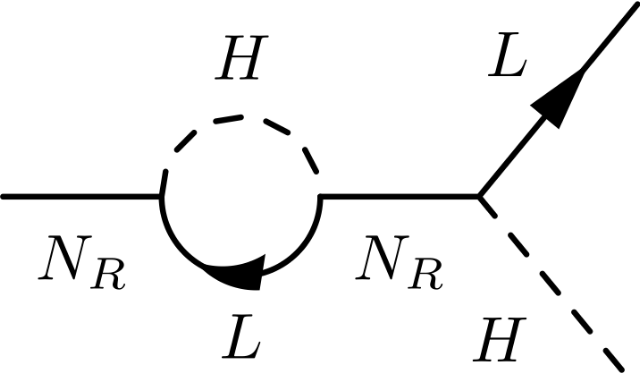} \includegraphics[width=4cm]{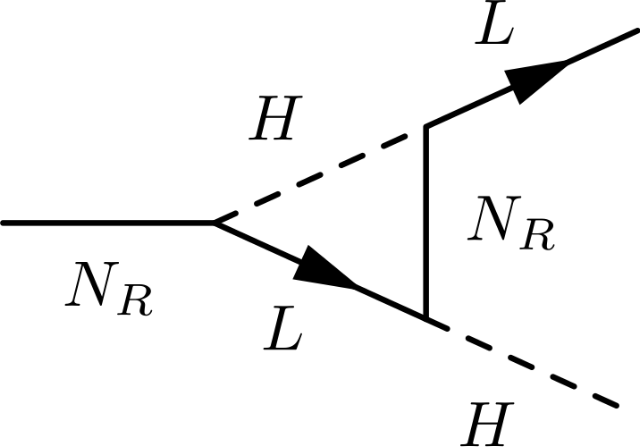}\\
  \caption{N Decay and CP violation }\label{fig1}
\end{figure}

CP violation is provided by the interference of tree-level and one loop diagrams, all controlled
by $\lambda$. Unless there is a special enhancement, we may thus expect the amount of CP violation to
be of order $\lambda^4$, while the direct decays are of order $\lambda^2$. The proportion of CP violating
decay for each heavy $N$ is thus expected to be of order $\lambda^2$ (see Fig \ref{fig1}).
Since other effects tend to dilute the baryogenesis effect, this amount of CP asymmetry must exceed the
wanted early universe asymmetry, namely $\epsilon > 10^{-8}$

The out-of-equilibrium condition states that the decay rate must be slower than the Universe expansion at
the time of decoupling (that is, roughly at temperature  $ T \approx m_N$).
Here $g^*$ is the effective number of degrees of freedom at that time.
\begin{eqnarray}
 \nonumber
  \Gamma &\simeq& \lambda^2 m \\
  \nonumber
  \Gamma &\ll& H \\
  H &=& \sqrt{g^*} T^2 /(10^{19} GeV )
\end{eqnarray}

We group in Table \ref{ordermagnitude} the various constraints on $\lambda$ and $m_N$,
adding the request to get reasonable light neutrino masses (say, of
order $0.01 eV$, through the see-saw formula $m_{\nu}= \lambda^2 v¨2 /M$,
where $v\approx 100 GeV$ is the electroweak symmetry breaking.

\begin{table}
  \caption{Bounds on $m_N$ (in GeV) for various $\lambda$ , assuming a light neutrino mass of order .01 eV}
  \label{ordermagnitude}\begin{tabular}{|c|c|c|c|}
  \hline
  $\lambda$ & Light neutrino mass  & Out of Equilibrium decay   & enough CP violation \\
 & $m_N \sim (GeV)$  & $m_N > (GeV)$  &  \\
  \hline

  $10^{-5}$ & $10^{7}$ & $10^{8}$ & needs tuning \\
  $10^{-4}$ & $10^{9}$ & $10^{10}$ & bordeline \\
  $10^{-3}$ & $10^{11}$& $10^{12}$ & yes \\
  $10^{-2}$& $10^{13}$ & $10^{14}$ & yes \\
  $10^{-1}$ & $10^{15}$ & $10^{16}$ & yes \\
  1 & $10^{17}$ & $10^{18}$ & yes \\

  \hline
\end{tabular}
\end{table}

As seen clearly from this table, the leptogenesis mechanism is fairly resilient over a wide range of
$\lambda, m_N$, but tuning becomes needed for low values of these parameters. Such tuning can take place
either through adjusting the individual elements of the Yukawa coupling matrix $\lambda$, or a considerable
enhancement can be found by making the self-energy diagram nearly resonant. This is obviously another kind
of tuning, which requests the  heavy neutrinos $N_1, N_2, ..$ to be nearly degenerate.
If the mass splitting is of order $\lambda^2$, the CP violation asymmetry can then be considerable.
Arguably, very low energy leptogenesis could then take place ~\cite{lowscale}.

\section{Improving or Falsifying Leptogenesis}

As announced, the main point of this note is to stress that, even if it is extremely difficult to
establish leptogenesis, it could at least be falsified. In particular, we contend that the observation
at present or future colliders (that is in practice in the TeV range) of $W_R$'s would make the canonical
form of leptogenesis (the case outlined above, with the lepton number carried by neutrinos) untenable.

The possible observation  of a $W_R$ will  of course be justification enough for its consideration! Still,
a few words of motivation for such a particle may be useful, and may help put back in context the whole
leptogenesis approach.

In my view indeed, introducing singlet fermions like the $N's$ of ad-hoc mass (quite separate from the
electroweak and grand unification scales) and Yukawa couplings,
if done outside a broader context, is mainly a reparametrization of an effective Lagrangian, and
involves no less fine tuning than putting by hand the small parameters this construction replaces. The situation is entirely different if such new particles are related to a wider (for instance, gauge) structure, in
which case a much more compelling picture emerges.

Without being specific about the wider gauge structure (one may think of SO(10), E6, or broader schemes),
some generators and their associated gauge bosons will typically involve $l_R - \nu_R$ (or in the present
notation $l_R -N$) transitions. They will also presumably couple to the right-handed quark structure.
For this reason, we consider specifically the case of $W_R$. Other effects may be associated with the other
members of the extended structure, notably extra $Z's$, or scalars, but we expect (at least in the
case of canonical leptogenesis considered here) that they will usually play in the same direction.

Including the $W_R$ sector was of course already considered, notably in \cite{Carlier:1999ac} and \cite{Cosme}.
In both cases, the study was involved with very heavy extra gauge bosons, and the way they would affect
leptogenesis and low-energy implications.
The most obvious result, as shown in \cite{Carlier:1999ac} is that the presence of $W_R$ will introduce new,
CP-conserving decay channels, potentially large, and lead to an extra dilution
of the generated lepton asymmetry, up to the point that the case $M_{W_R} < m_N$ is virtually excluded.
This is however by far not the only effect. Further reduction of leptogenesis is associated to diffusion
processes, but quite interestingly , the opposite effect may also arise.

As shown indeed in \cite{Cosme}, the presence of $W_R$ may play a determinant role when the $N$ population
has been destroyed through inflation and needs to be rebuilt. If, as sometimes assumed, the $N$ don't couple
directly to the reheating process, small Yukawa couplings (associated to particularly light neutrinos) would in fact preclude
the rebuilding of a sufficient population. In that case, the presence of right-handed gauge interactions
saves the day, and destroys the possible lower limits on neutrino masses which could be induced.

\section{The main effects}
We start thus by including the new interaction term
\begin{equation}
{\cal L}_{W_R} =  \frac{g}{\sqrt{2}}
W_R^{\mu} \left( \bar{u}_R \gamma_{\mu} d_R + \bar{N} \gamma_{\mu} \,l_R \right)
\end{equation}

The most evident effect is on the decay channels. Since these  are CP-conserving, they
introduce a dilution of the asymmetry $\epsilon^{(0)}$ generated in the standard case:

\begin{equation}
\label{dilution}
\epsilon = \frac{\Gamma_N ^{(l)} - \overline{\Gamma}_N ^{(l)}}{\Gamma_{tot}^{(l)} +
\Gamma_{tot}^{(W_R)}} \equiv \epsilon^{(0)} \frac{\Gamma_{tot}^{(l)}}{\Gamma_{tot}^{(l)} +
\Gamma_{tot}^{(W_R)}}
\end{equation}

We denote the abundances $Y_i\equiv n_i/s$, $Y_{\cal B}\equiv Y_{B}-Y_{\bar{B}}$,
$Y_{\cal L}\equiv Y_L-Y_{\bar{L}}$, where $n_i$ the comoving number
density of the species "i", "eq" refering to the equilibrium number
density,  and $s$ the comoving entropy density.
In a now standard notation,
\begin{equation}
Y_{\cal B} = Y_{\cal L} \, r_{{\cal L}\rightarrow {\cal B}} = \varepsilon_N \,\eta \, Y_N^{eq}(T\gg m_N) \,
 r_{{\cal L}\rightarrow {\cal B}}.
\end{equation}

where $r_{{\cal L}\rightarrow {\cal B}}$ is the conversion rate of lepton to baryon number at the electroweak
phase transition, and $\eta$ is referred to as the efficiency, and involves all the effects of evolution
of the lepton number under the Boltzmann equations.

To facilitate the discussion, we will now slightly depart from the usual conventions, and will include
the above-mentioned dilution effect ( that is , the factor  $\frac{\Gamma_{tot}^{(l)}}{\Gamma_{tot}^{(l)} +
\Gamma_{tot}^{(W_R)}}$ appearting in eq. \ref{dilution}) in the expression of the efficiency $\eta$.
Using this convention, $\epsilon \rightarrow \epsilon^{0}$.

We now set to examine if the dilutions effects due to a light $W_R$ are sufficient to make canonical
leptogenesis impossible. For this purpose, we can, in the above convention, replace $\epsilon$ by the
largest possible value. While both degenerate and non-degenerate cases are considered in \cite{Frere:2008ct}, we will consider here
the least favorable situation (for our purpose of disproving the mechanism), namely $\epsilon =1$ (thus
allowing for resonance enhancement).

A very important effect arises from the scatterings. Indeed, the $W_R$ have the important property
of interacting with gauge strength with the right-handed quarks in the thermal plasma. This keeps them
in thermal equilibrium, but also enhances the effect of the scatterings, since the "relic" $N$ particles
interact through $W_R$ with normally abundant quarks and light leptons (at the difference of the case where
the relic particles must annihilate mutually).

\begin{figure}

  \includegraphics[width=6cm]{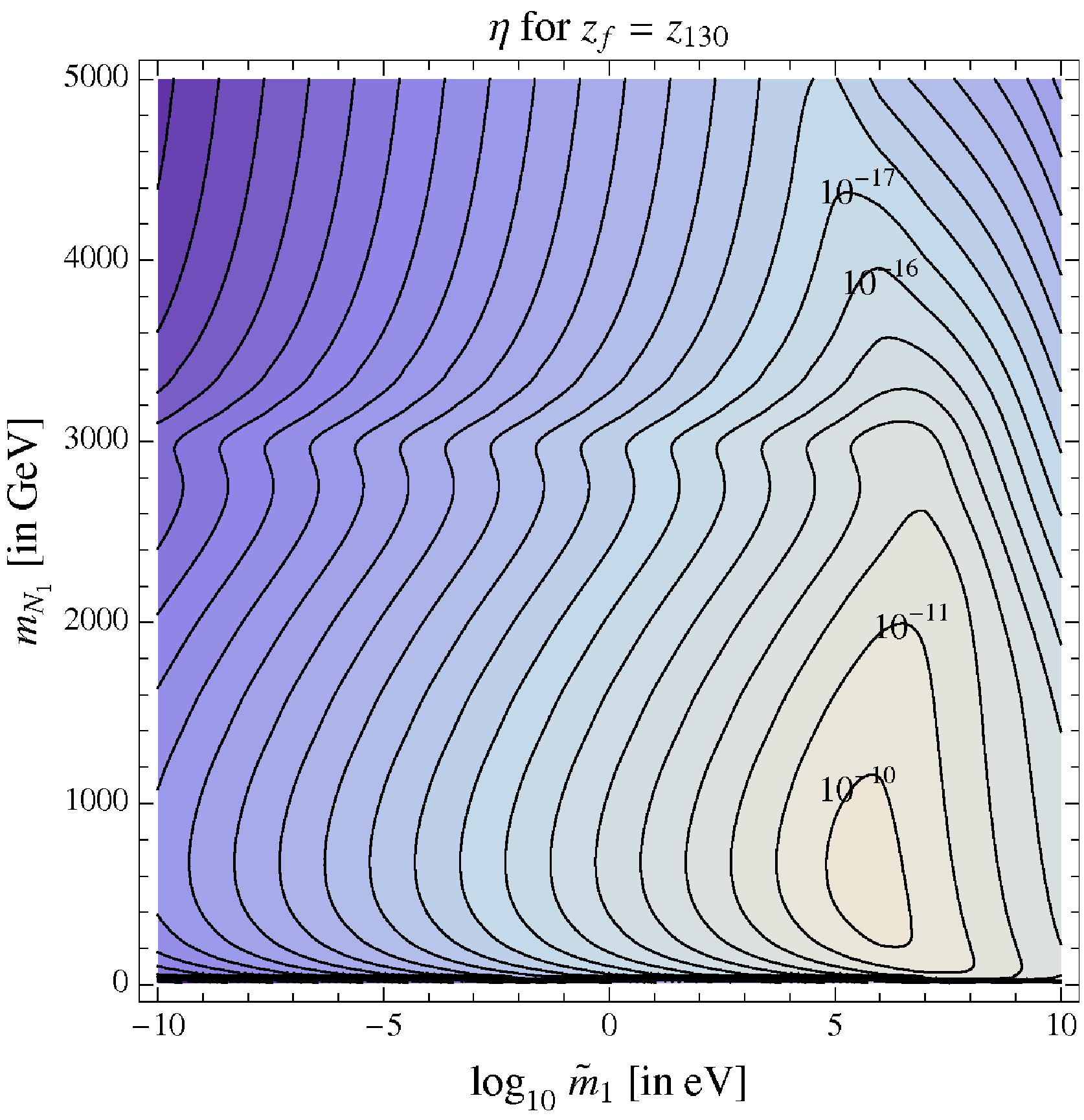}\includegraphics[width=6cm]{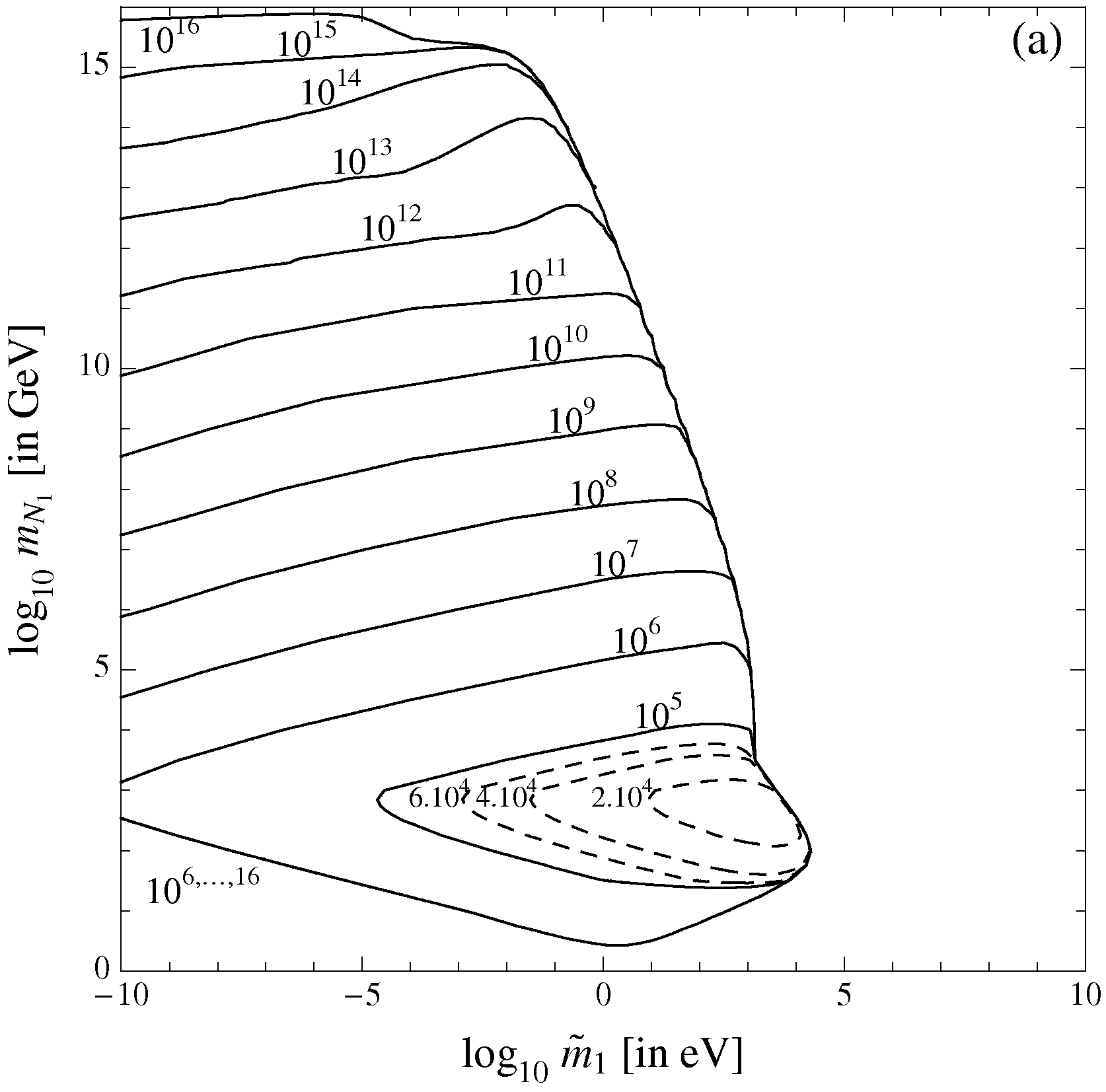}\\
  \caption{Main results: the panel at the left gives for $M_{W_R}=3TeV$ the efficiencies reached
  as a function of  $m_N$ and $\tilde{m_1}=v^2{\lambda_\nu^\dagger \lambda_\nu}_{11}/{m_N}$, ($z-f$ refers to the scale at which the decoupling
  of the sphaleron conversion mechanism is assumed). The panel at the right gives the lower limit
  acceptable for the $W_R$ mass (in GeV) assuming a maximal leptonic
  asymmetry due to CP violation ($\epsilon =1$) \label{efficiencies}}
\end{figure}

The results are most easily read from Fig. \ref{efficiencies}, were we give (in the right-most panel), the
lower bound on $M_{W_R}$ compatible with leptogenesis. The values, given in GeV are clearly out of reach of
the currently operating or planned colliders (we find a lower bound of 18 TeV in the present case).
  As an example, we also list (in the left-most panel) the actual
efficiencies (which would also be the lepton number generated, in case $\epsilon = 1$) for $M_{W_R}= 3 ~TeV$,
a value reachable at the LHC. (remember that a leptonic excess of at least $10^{-8}$ must be generated to
accomodate the currently observed matter asymmetry.

The above considerations put some new urgency to the quest at colliders for $W_R$ bosons,
or possibly even light $N$.  In particular, the search \cite{Bayatian:2006zz} should be extended to include the situation where the $N$ is heavier
than the $W_R$, a case where the exclusion of leptogenesis is even more severe.

\section{acknowledgments}
This work was performed in collaboration with Thomas Hambye and Gilles Vertongen,
with support from IISN and Belgian Science Policy (PAI VI/11)

\end{document}